\documentstyle[12pt,aaspp4,psfig]{article}
\lefthead{Hashimoto \& Oemler.}
\righthead{Concentration-Density Relation in LCRS}

\begin{document}

\title{THE CONCENTRATION-DENSITY RELATION OF GALAXIES IN LAS CAMPANAS REDSHIFT SURVEY}

\author{Yasuhiro Hashimoto\altaffilmark{1,2}
   and Augustus Oemler, Jr.\altaffilmark{1,2}
   }

~~\\
~~\\
\centerline{\em Accepted for publication in Astrophysical Journal}

\altaffiltext{1}{Department of Astronomy, Yale University, P.O. Box 
    208101, New Haven, CT 06520-8101; hashimot@astro.yale.edu,oemler@astro.yale.edu}
\altaffiltext{2}
{Carnegie Observatories, 813 Santa Barbara St.,
    Pasadena, CA 91101 USA; hashimot@ociw.edu,oemler@ociw.edu}
\begin{abstract}

We report the results of the evaluation of the
``concentration-density'' relation of galaxies in the local universe,
taking advantage of the very large and homogeneous data set available
from the Las Campanas Redshift Survey (\cite{sh96}). This data set
consists of galaxies inhabiting the entire range of galactic
environments, from the sparsest field to the densest clusters, thus
allowing us to study environmental variations without combining
multiple data sets with inhomogeneous characteristics. Concentration
is quantified by the automatically-measured concentration index $C$,
which is a good measure of a galaxy's bulge-to-disk ratio. The
environment of the sample galaxies is characterized {\em both} by the 
three-space local galaxy density and by membership in groups and clusters.

We find that the distribution of $C$ in galaxy populations varies both
with local density and with cluster/group membership: the fraction of
centrally--concentrated galaxies increases with local galaxy density,
and is higher in clusters than in the field. A comparison of the
concentration--local density relation in clusters and the field shows
that the two connect rather smoothly at the intermediate density
regime, implying that the apparent cluster/field difference is only a
manifestation of the variation with the local density.  We conclude
that the structure of galaxies is predominantly influenced by the
local density and not by the broader environments characterized by
cluster/field memberships.

\end{abstract}


\section{INTRODUCTION}

It has long been realized that there are differences between
populations of galaxies inside and outside of clusters (e.g. Hubble \&
Humason 1931). Quantitative analyses (Oemler 1974; Dressler 1980) have
shown that the fractions of elliptical and S0 galaxies rise in the
cores of clusters compared to their peripheries, and several studies
have demonstrated that this environmental dependence extends to less
dense environments (e.g. \cite{bh81}; \cite{ds82}; \cite{pg84};
\cite{gh86}).

These classical investigations of the dependence of morphology on
environment have some weaknesses. Firstly, these studies have often
been based on small samples (on the order of 10$^2\sim 10^3$
galaxies). With samples of this size, it is difficult to make a
statistically sound comparison, particularly after binning galaxies by
different environments.  Moreover, these samples are often composed of
multiple data sets with heterogeneous characteristics, such as
different Hubble classifications by different observers, varying image
quality, and even different selection criteria of sample objects, all
of which can cause spurious results if much care is not taken for their
inhomogeneities. However, the most serious
consequence of small samples is their inability to discriminate
between effects dependent on the very local environment, as
characterized, for example, by local galaxy density, from the effects
of larger-scale environments, such as membership in a cluster.
The significance of the distinction between local and broader
environments has been a matter of much contention in cluster studies
(see, for example, Dressler 1980 vs.  Whitmore \& Gilmore 1991).

Moreover, almost all previous studies of the morphology-environment
relation have used Hubble type as a measure of galaxy properties. This
was both natural- Hubble type is easy to determine- and
effective. However Hubble type is a compound of two galaxy
characteristics: star formation rate and bulge-to-disk ratio (B/D). Although
the success of the Hubble System demonstrates that these two
parameters are well-correlated along the Hubble Sequence, it cannot be
taken for granted that the unknown processes by which environment
effects morphology would act equally on these two very different
characteristics.

That star formation is not perfectly correlated with B/D
was pointed out some time ago by van den Bergh (1976), who
introduced the term ``anemic'' to refer to galaxies whose star
formation rate is lower than typical for galaxies of their Hubble
types. In recent years many studies have sought, and some have found,
evidence that star formation rates in galaxies of a given Hubble type
differ in different environments (e.g. Gisler 1978;
Kennicutt 1983;  Kennicutt et al.
1984; Gavazzi \& Jaffe 1985; Moss \& Whittle 1993). In a previous paper
(Hashimoto et al. 1998, hereafter Paper I), we examined this question
using a sample of 15749 galaxies drawn from the Las Campanas Redshift
Survey (Shectman et al. 1996), and found that the star formation rate in
galaxies of a given bulge-to-disk ratio varies significantly with
environment, in a complex manner.

Clearly, then, the morphology-environment relation is a compound
effect, the product of a variation with environment of the dependence
of star formation rate on B/D, and a (possible)
variation with environment in the mix of galactic bulge-to-disk ratios.
Having established the former in Paper I, we turn in this
paper to the latter effect.  In this paper, we report the results of the 
evaluation of
the ``concentration -- density'' relation for galaxies in the local
universe, derived from the very large and homogeneous data set in the
Las Campanas Redshift Survey. 
Automatically measured concentration index (Okamura, Kodaira, \& Watanabe 1984),
which is more closely related to the B/D than Hubble type,  
was used to quantify the light distribution of galaxies.
The concentration index not only suits our need for relatively ``star formation
free'' measure of galactic structure, it
is also more robust against image degradation, and  easier to 
measure automatically. It is, thus, ideal for a large galaxy survey 
such as the LCRS, where the sample size is $\sim$ 10$^4$ and most of
the galaxies consist of on the order of 10$^2$ resolution elements.
The Las Campanas Redshift Survey consists of galaxies
inhabiting the entire range of galactic environments, from the
sparsest field to the densest clusters, thus allowing us to study
environmental variations without combing multiple data sets with
inhomogeneous characteristics. Furthermore, we can also extend our
research to a ``general'' environmental investigation by, for the
first time, decoupling the local galaxy density from the membership in
associations.

\section{DATA}
Here we briefly describe our survey parameters; the reader is referred
to Shectman et al.\ (1996) for further details.  The LCRS consists of
26418 galaxies, with a mean redshift $z = 0.1$, and a depth of about
$z=0.2$. The survey galaxies were selected, using isophotal
and central magnitude criteria, from a CCD-based catalog obtained in a
``hybrid'' Kron-Cousin $R$ band. The survey covers over 700 square
degrees in six $1.5^{\circ} \times 80^{\circ}$ ``slices''.  The first
20\% of the redshifts were obtained using a 50-object fiber-optic
spectrograph. The nominal isophotal magnitude limits were $16.0 \leq R
\leq 17.3$, and an additional central magnitude limit excluded the
lowest 20\% of galaxies by central surface brightness.  The rest of
the redshifts were obtained with a 112-object fiber system, with
isophotal limits of $15.0 \leq R \leq $17.7, and exclusion of the
lowest 5-10\% of galaxies by central surface brightness.

CCD images were taken at the 1m Swope Telescope at Las Campanas using
the drift scan technique through a Thuan-Gunn $r$ filter. The exposure
time depends on the size of the CCD chip and the cosine of the
declination, but is typically near 1 minute. Three CCDs were used
over the course of this survey. The first was a thinned 800 $\times$
800 TI chip, which was used with a focal reducer to obtain 1\farcs05
pixels.  This was replaced in March 1990 with a thick 2K $\times$ 2K
LORAL chip, without a focal reducer and binned 2 $\times$ 2 to produce
0\farcs865 pixels. The third CCD, introduced in 1992, is a thick 2K
$\times$ 2K Tektronix chip used unbinned without a reimaging optics to
provide 0\farcs692 pixels.
 
\section{STRUCTURAL PARAMETERS}
 
Morgan (1958) first proposed a classification of galaxies based on the
central light concentration, while Bothun (1982) recommended the use
of visual bulge-to-disk ratio as a more fundamental basis of galaxy comparison.  These
``visual'' quantities have recently been superseded by the
automatically determined concentration index $C$ (Okamura et al. 1984;
Doi, Fukugita, \& Okamura 1993; Abraham et al. 1994).
Concentration retains a very close connection
with estimates of bulge-to-disk ratios, thus it is a purer measure
of one of the two physical parameters determining Hubble type. 
It is also quite robust against
image degradation and much easier to measure automatically than
the bulge-to-disk ratio itself, thus ideal for morphological
classification of a large galaxy survey, such as the LCRS, where the
sample size is $\sim$ 10$^4$ and most of the galaxies consist of
$\sim$ 10$^2$ resolution elements.  The concentration index measures
the intensity weighted second moment of the galaxies and compares the
flux between the inner ($r$ $<$\ 0.3) and outer ($r$ $<$\ 1) isophote
to indicate the degree of light concentration in the galaxy images.
Here $r$ is a normalized radius which is constant on an elliptical
isophote and it is normalized in such a way that $r$ is unity when the
area within the ellipse is equal to the detection area of a galaxy.
(For further details of the definition, please see Hashimoto et
al. 1998.)

Another subsidiary parameter $A$ (Asymmetry Index) is also measured for 
LCRS galaxies for the purpose of excluding morphologically
peculiar galaxies which are expected to introduced some scatter
in the relationship between $C$ and B/D.
$A$ is measured by first rotating a galaxy image by 180 degrees around the 
image center, then subtracting the rotated image from the original unrotated
one. The residual signals above zero are summed and then normalized
by flux above the sky to calculate $A$.
Note that merged objects are deblended by the detection algorithm 
SExtractor (Bertin 1994, hereafter SXR)
and  $A$ is measured for the signal inside the 
the outer isophote of the deblended central component. 
For highly blended objects where SXR cannot deblend the merging components,
$A$ is measured for the whole blended component.

\section{ENVIRONMENTAL PARAMETERS}
We characterize 
the environment of LCRS galaxies by calculating the
local galaxy density $\rho$ around each 26418 galaxy using a nearest
neighbor technique. 
Namely, given a particular galaxy, we take the local galaxy density to
be $\rho=3/V$, where $V=\frac{4}{3}\pi D^{3}$ and $D$ is the 
three-dimensional redshift-space distance from the
given galaxy to its third nearest neighbor.
To account for 
the effect of the variation of the survey selection function
at different redshifts,
we replace the galaxy count by
$w(z_{j})W_{j}$, so that the 
local galaxy density $\rho$
around a galaxy $i$ becomes
$\rho_{i}=\Sigma ^{3}_{j=1}w(z_{j})W_{j}/V,$
where $j$ represents the rank of the nearest neighbors from galaxy $i$.
Here, $w(z_{j})$ is a weight
\begin{equation}
 w(z_{i}) =
   \int^{M_{2}}_{M_{1}} \phi(M)dM
 \left/
 \int^{min[M_{max}(z_{i}),M_{2}]}_{max[M_{min}(z_{i}),M_{1}]}\phi(M)dM
 \right. \ ,
\end{equation}
where $M_{1}, M_{2}$ are the absolute magnitude limits in which we are
interested, and
$ M_{max}(z_{i})$ and $ M_{min}(z_{i})$ are
the absolute magnitude limits, at the redshift of galaxy $i$,
corresponding to the apparent magnitude limits for the field
containing galaxy $i$.
We describe the differential luminosity
function $\phi$ by a Schechter function (Schechter 1976) 
with parameters $\phi^{*} =
0.019 \ h^3$~Mpc$^{-3}$, $M_{R}^{*} = -20.29 + 5 \log h,$ and $ \alpha
= -0.70 $ (Lin et al. 1996), which we assume to be invariant with
redshift.
In addition, another weight $W_i$ is calculated for each galaxy $i$ to
take account of the field-to-field spectroscopic sampling variations.
The spectroscopic completeness of
a field decreases as the projected density of galaxies
in the field increases, since each spectroscopic field is observed
only once, using a maximum of 50 or 112 fibers.
This effect is corrected by setting
$ W_{i}$ to be the inverse of the fraction of spectroscopically
observed galaxies in the field containing galaxy $i$.

As the second environmental parameter, cluster or rich-group membership
is determined for each galaxy in the LCRS,
using the 
three-dimensional ``friends-of-friends'' group identification algorithm
(Huchra \& Geller 1982). The algorithm finds all pairs within a projected
separation $D_{L}$, and within a line of sight velocity difference $V_{L}$.
Pairs with a member in common are linked into a single group.
This linking makes the membership more sensitive to the environment on larger
scales than the scale of the local density parameter. 
The selection
parameters $D_{L}$ and $V_{L}$ are scaled to account for the magnitude
limit of the LCRS survey, and defined as
$D_{L}=S_{L}D_{0}$\ and  $ V_{L}=S_{L}V_{0}$.
Here the linking scale S$_{L}$ is calculated by
$S_{L} = [\rho^{'}(d_{f})/\rho^{'}(d)]^{1/3}$,
where $\rho^{'}(d)$ is the galaxy number density, at the mean comoving
distance $d$ of the galaxy pair in question, for a
homogeneous sample that has the same selection function as
the LCRS.
The distance $d_{f}$ is the fiducial
comoving distance at redshift z$_{f}$ (we chose z$_{f}$=0.1) at
which we define $D_{0}$ and
$V_{0}$. The density enhancement contour surrounding
each group is related to $D_{0}$ by
$\Delta\rho/\rho = [3/4{\pi}D_{0}^{3}\rho^{'}(d_{f})]-1.$
The values of $D_{0}$ (or $\Delta\rho/\rho$) and $V_{0}$  used are taken
from the LCRS group catalog (c.f. Tucker 1994), and are 
$D_{0}$= 0.72
$h^{-1}$ Mpc (or $\Delta\rho/\rho$ =80) and $V_{0}$ = 500 km/s, which
are determined by several semi-quantitative constraints similar to
those used in
Huchra \& Geller (1982). 
to avoid biasing the velocity dispersions of groups. 
The environmental parameters are further discussed in Hashimoto et. al (1998).

\section{RESULTS}
\subsection{Comparison of Image Parameter to Hubble Type}
Hubble type classification, completely independent from $C$ measurements,
was performed visually by one of authors (A. O.) for $\sim$ 200
LCRS galaxies. These galaxies are selected semi-randomly from three
typical LCRS $1.5^{\circ} \times 80^{\circ}$ slices, so that the
sample covers a wide variety of morphology and galaxy environments.
For galaxies with disturbed morphologies, a flag
is assigned, as well as best estimated Hubble types.
An additional flag is assigned by A. O. for galaxies with uncertain or
indeterminate morphologies, which are mostly galaxies with extremely
small angular size ($\leq$ 55 pixels).
Fig. 1 shows the relationship between the concentration index $C$ and
A. O. classification. Galaxies 
with angular size less than equal to 55 pixels are excluded, as are
galaxies with disturbed or uncertain morphologies.
Fig. 1 confirms an overall correlation between $C$ and Hubble type,
where early Hubble type galaxies show higher $C$ values compared to late
Hubble type galaxies.
In addition, Fig. 1 shows that
earlier type galaxies are well separated from later type galaxies,
while subtle distinctions, such as distinguishing ellipticals from S0,
are not effectively made. Also note that a clearer separation can be made 
between Sa \& Sc, rather than the conventional early/late type separation 
at a borderline between S0 \& Sa.

\subsection{Comparison between Clusters and the Field}

Fig. 2 shows the distribution of $C$ for cluster and
field galaxies.
The ``cluster'' sample consists of ``cluster or rich group'' galaxies defined 
in \S4.
Meanwhile, galaxies outside of clusters, hereafter ``field galaxies'',
are identified by removing cluster galaxies
from the entire sample, except that this time,
a lower $\Delta\rho/\rho$=40 (instead of $\Delta\rho/\rho$=80) contour 
is used to ensure that galaxies
in the outskirts of clusters are excluded from the field sample.
The solid line represents cluster galaxies,
while the dotted-dashed line represents galaxies in the field.
Galaxies with extremely small angular size ($\leq$ 55 pixels), as well as 
apparent disturbed galaxies ($A > $0.09) are excluded.
The $C$ distribution of the cluster galaxies is skewed toward higher
B/D (bigger $C$).
This result is qualitatively  consistent with the morphology-density
relation using the standard Hubble type, however,
the ``concentration-density relationship'' in Fig. 2  
is relatively free from the systematic and random Hubble type error
caused by the image degradation and the star formation activities of galaxies.

We can also examine the distribution of C as a function of the richness
of the clusters.
Table 1 lists the percentiles of the various C classes in various
cluster subdivisions.
Cluster and field
definitions are the same as Fig. 2.
Additionally, rich and poor cluster subsets are introduced, as
defined by their total luminosities
$L_{T}$ (Tucker 1994).
Rich clusters are defined as clusters with
$L_{T}\ \geq\ 5\times10^{11}L_{\odot} $, while poor clusters are clusters
with
$L_{T} \leq\ 0.5\times10^{11}L_{\odot}$.
Three C classes are defined as, $C=$0.35-0.50, 0.25-0.35,
and 0.10-0.25.
The numbers in parentheses are
the total number of galaxies in each environmental class.
Table 1 reproduces the general trend that cluster galaxies have
higher $C$ compared to the field galaxies.
When you compare the rich and poor cluster galaxies,
there is a similar but more
distinctive trend
that galaxies in rich clusters have higher $C$ with respect to galaxies in
poor clusters.
These two (cluster-field and rich-poor) comparisons seem to be consistent
with each other in that galaxies in rich environments
tend to have larger bulge-to-disk ratios (higher $C$). However,
remarkable are the percentiles in the poor clusters.
Poor cluster galaxies seem to show {\it lower} $C$ with respect
to the field galaxies.
This might suggest that
galaxies in denser regimes have smaller B/Ds, which seems to
contradict the implication from the previous two comparisons.
Note, however, that the trend is not statistically conclusive;
a $\chi^{2}$\ test shows that the two proportions are
significantly different at significance level $\alpha$\ = 7$\times10^{-1}$,
4$\times10^{-1}$, and 2$\times10^{-1}$,
for $C=$ 0.35-0.50, 0.25-0.35, and 0.10-0.20, respectively.
Further investigations with even larger galaxy samples are encouraged to 
test this possibility.

\subsection{Correlation with Local Density}
Fig. 2 shows the dependence of the galactic concentration on
the membership in associations. 
We can also examine the dependence of C on the local density.
Fig. 3 shows the population ratio of late to early type, as a function
of the local 3-space density for LCRS galaxies. 
Bars are root N error.
Early and late type are determined using $C$; early type ($C >$0.27) and
late type ($C \leq$ 0.27). Again, galaxies with extremely small angular size 
($ \leq$ 55 pixels), as well as apparent disturbed galaxies
($A > $0.09) are excluded.
Fig. 3 shows that the population ratio of late to early type galaxies 
correlates with local galaxy density, in such a way that late type
galaxies tend to be more abundant in less dense environments.

Fig. 2 and 3 together show that the concentration of galaxies 
is correlated with both the local density and 
membership in associations.
Reviewing Fig. 2 \& 3, one might be tempted
to conclude that the cluster/field difference (Fig. 2) is simply
a manifestation of the variation with the local density. Conversely, 
Fig. 3 might merely represent a superposition of different membership
dependencies.
To understand the mechanism of the environmental influence on
the concentration of galaxies, it is important to further investigate
the behavior of the correlation in Fig. 3 inside and outside
of clusters separately.

\subsection{Density Dependencies Inside/Outside Clusters}

Fig. 4a shows the same relation as Fig. 3 for the cluster
sample, alone.
Overall, Fig. 4a shows a qualitatively similar correlation as Fig. 3;
the late type galaxies tend to be more abundant in less dense regions. 
The fact that the ratio correlates with the local density inside clusters 
suggests that the galaxy concentration might be influenced by the 
environment of a relatively localized volume surrounding a galaxy, rather 
than the volume comparable to the size of a entire cluster.
However, it is still possible that 
the correlation in Fig. 4a might be due to the local 
environments {\em specific} to cluster environments (namely the local density
here is not of universal nature and we do not see this local density
dependence in the field), or the local density is universal but 
the cluster environments might 
have independent influence on the concentration of galaxies.
To find out this, it is necessary to 
to further investigate the behavior of the correlation outside of
clusters.

Fig. 4b shows the identical population ratio to Fig. 4a, but now for the
{\it field}\ sample alone.
Overall, Fig. 4b shows qualitatively the same relation as Fig. 4a,
namely,  
the concentration of galaxies is correlated with the local density
even outside of the clusters.
This implies that the dependence on the local density is universal 
and the concentration of the galaxies is perhaps predominantly 
influenced by the local density.
The superposition of Fig. 4a and 4b further reveals that the two
relationships connect rather smoothly (the inset), confirming that the cluster/field
difference (Fig. 2) is perhaps only a manifestation of the variation with 
the local
density, thus we conclude that the morphology of galaxies is
predominantly influenced by the local density and not by the
broader environments characterized by the cluster/field memberships.

\section{SUMMARY}

The results of this study can be summarized as follows:

(1) The cluster/field subdivision shows that distributions of the
concentration index of cluster galaxies is skewed toward higher
bulge-to-disk ratios (bigger $C$). This result is qualitatively
consistent with the morphology-density relation with the standard
Hubble type.

(2) The central concentration of galaxies is also found to be
correlated with the 3-space local density, in such a way that late
type galaxies tend to be more abundant in the low local density
environments.

(3) Further investigation of the behavior of the local
density correlations inside and outside of clusters shows that the
cluster relation and field relation connect rather smoothly at the
intermediate densities, implying that the apparent cluster/field
difference is perhaps only a manifestation of the variation with the
local density.  

It thus appears that the bulge-to-disk ratios of galaxies are
predominantly influenced by the local density and not by the broader
environments characterized by cluster/field memberships. This result
is in striking contrast to the environmental dependence of star
formation rates explored in Hashimoto et al. 1998 (Paper I), which exhibit 
as great a
sensitivity to cluster/group membership as they do to local
density. In Paper I, we concluded that at least {\em two} processes
were at work by which environment influences star formation rates: one
which suppresses star formation in rich clusters, and another which
precipitates starbursts in intermediate richness environments. The
differing sensitivity of $C$ to environment strongly suggests that
structure is determined by yet a {\em third} process.

Unfortunately, the observed dependence of concentration on environment
provides little guidance about the nature of that process. Presumably,
galactic structure, unlike galactic star formation rates, is the
product of processes operating over the entire life of the galaxy, and
is not subject to short-term fluctuations. This is not inconsistent
with a dependence on local density, because the local densities are
correlated over very long timescales, except in the well-mixed cores
of rich clusters. Thus, any process sensitive to local conditions, and
operating over extended periods, could produce correlations like those
seen in Fig. 3. Local environment at the time of formation might
produce Fig. 3, but so might the integrated merger rate over the
life of the galaxy. Fig. 3 cannot be used to distinguish between
these alternatives, but the evolution with redshift of this relation
can, and such an observation provides the best hope of elucidating the
mechanisms responsible for determining galactic structure.

\acknowledgments
We thank referee for very careful reading
and comments which improved the paper.
This work was partially supported by NSF grant AST91-15446. The Las
Campanas Redshift Survey was supported by NSF grants AST87-17207,
AST89-21326, and AST92-20460. YH was partially supported by a Carnegie
Predoctoral Fellowship.

\clearpage
\begin{figure}
\centerline{
\psfig{figure=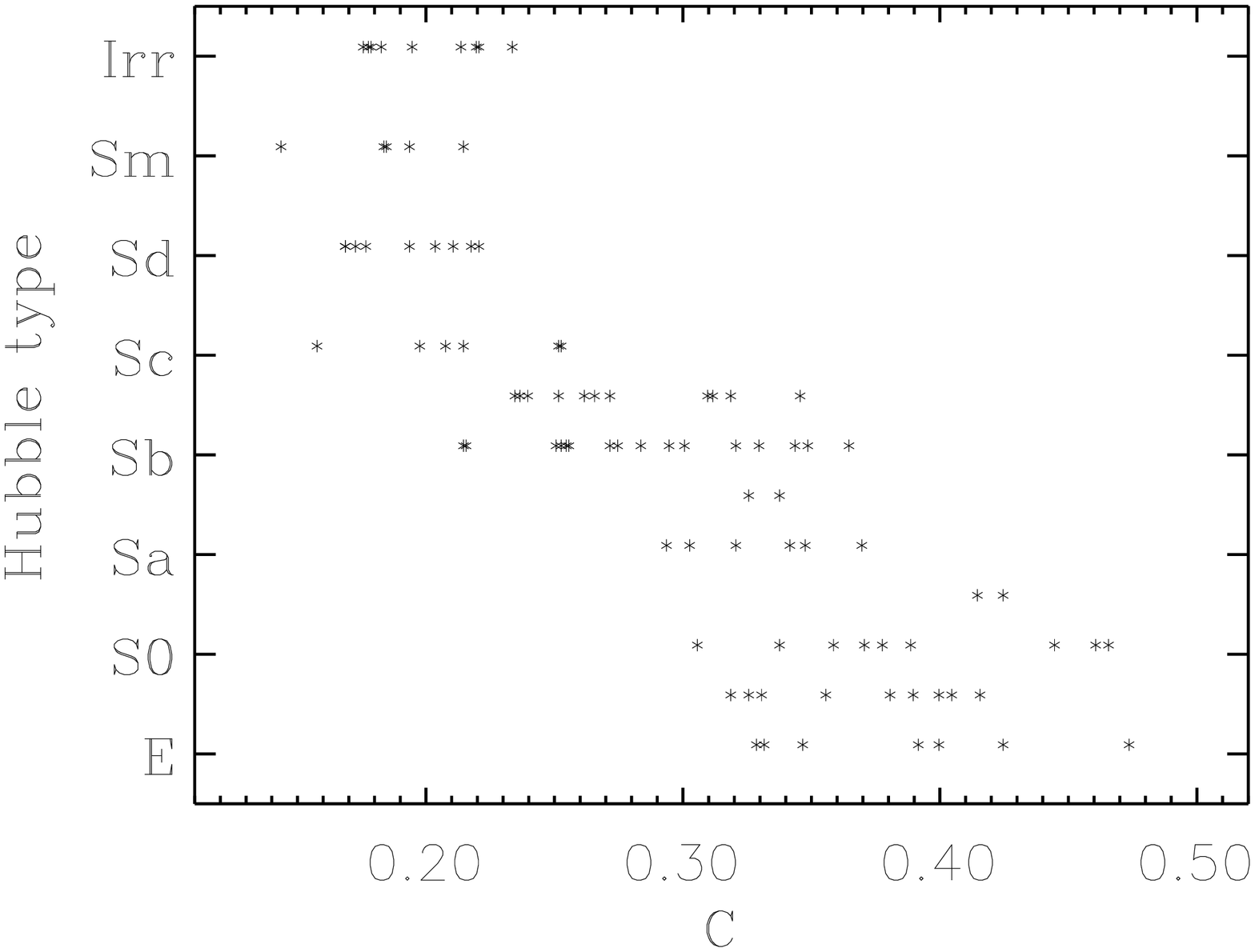,height=14cm,width=19.0cm,angle=90}
}
\figcaption[fig1.ps]{
 The relationship between automatically-measured concentration index
$C$ and Hubble classification by one of authors (Oemler) for $\sim$ 200
galaxies taken from the Las Campanas Redshift Survey (LCRS).
\label{fig1}}
\end{figure}

\begin{figure}[t]
\centerline{
\psfig{figure=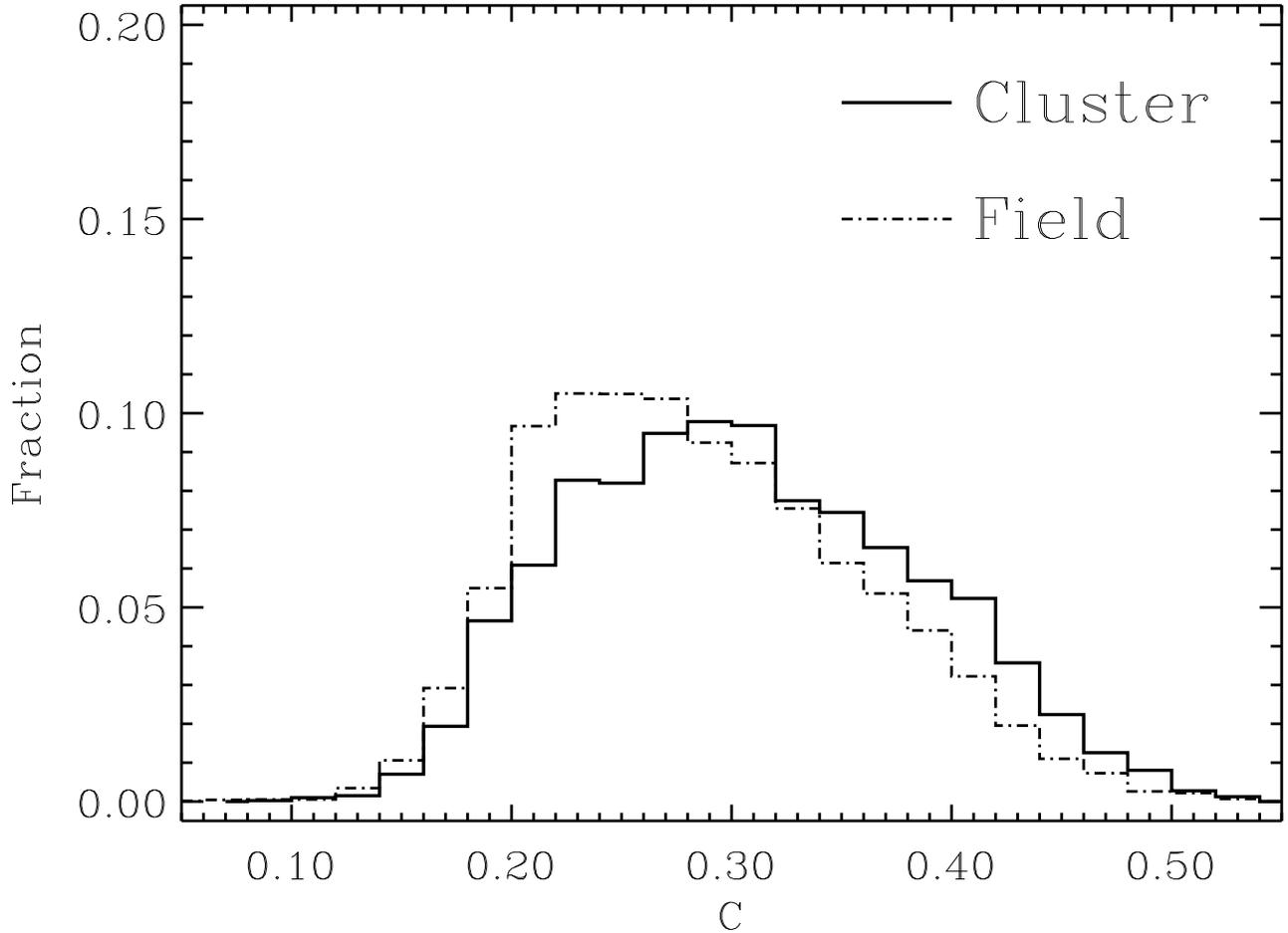,height=14cm,width=19.0cm,angle=90}
}

\figcaption[fig2.ps]{
The distribution of the concentration index $C$ for cluster and field
galaxies taken from the LCRS.
The cluster membership is determined using three-dimensional 
``friends-of-friends'' algorithm (Huchra \& Geller 1982).
The $C$ distribution of the cluster galaxies is skewed toward early type
(bigger $C$).
This result is qualitatively  consistent with the morphology-density
relation using the standard Hubble type, however,
the ``concentration-density relationship'' in this figure
is relatively free from the systematic and random Hubble type error
caused by the image degradation and the star formation activities of galaxies.
\label{fig2}}
\end{figure}

\begin{figure}
\centerline{
\psfig{figure=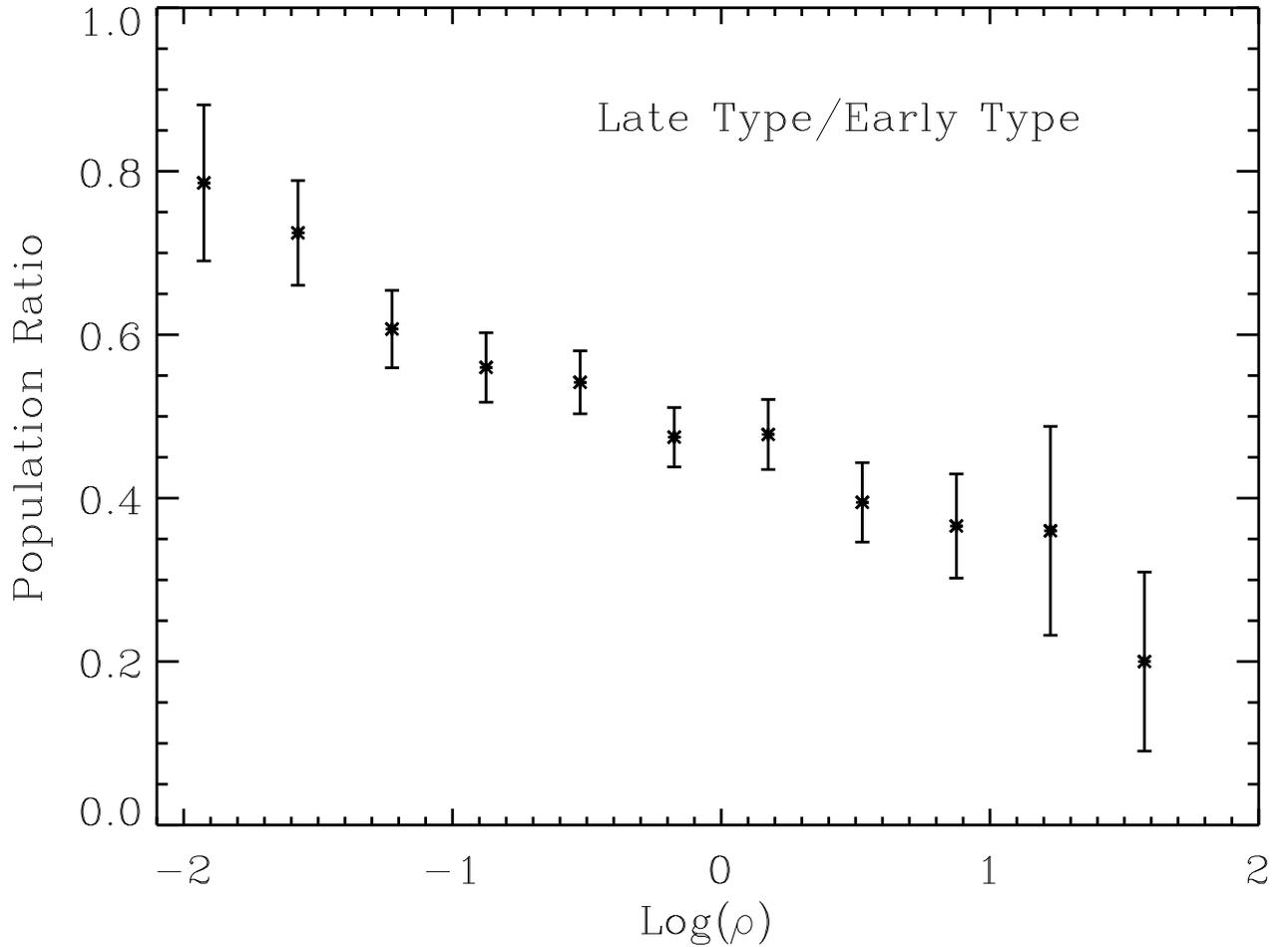,height=14cm,width=19.0cm,angle=90}
}

\figcaption[fig3.ps]{
The population ratio of late to early type, as a function
of the local 3-space density for LCRS galaxies.
Bars are root N error.
Early and late type are determined using $C$; early type ($C >$0.27) and
late type ($C \leq$ 0.27).
Fig. 2 and 3 together show that the concentration of galaxies
is apparently correlated {\em both} with the local density and the
memberships in associations.
\label{fig3}}
\end{figure}

\begin{figure}
\centerline{
\psfig{figure=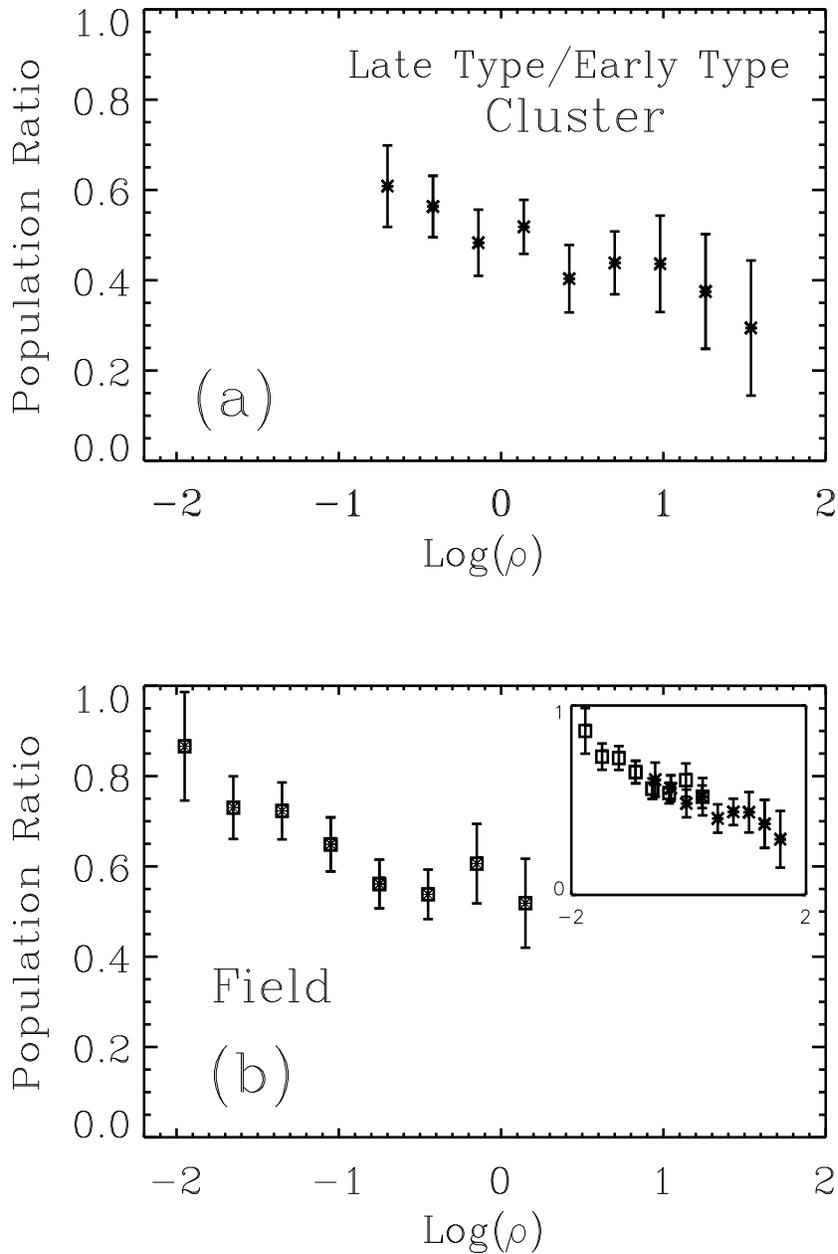,height=18cm,width=13.0cm}
}

\figcaption[fig4.ps]{
Fig. 4a and 4b show the same ($C $ vs. local density) relation as Fig. 3 for 
the cluster and field sample, respectively.
These figures show that the concentration of galaxies is correlated with the 
local density both inside and outside of the clusters, implying that
the dependence on the local density is universal
and the concentration of the galaxies is predominantly
influenced by the local density.
The superposition of Fig. 4a and 4b further reveals that the two
relationships connect rather smoothly (the inset), confirming that the 
cluster/field
difference (Fig. 2) is only a manifestation of the variation with
the local
density. We conclude that the concentration of galaxies is
predominantly influenced by the local density and not by the
broader environments characterized by the cluster/field memberships.
\label{fig4}}
\end{figure}

\clearpage

\begin{deluxetable}{crrrrrrrrrrr}
\footnotesize
\tablenum{1}
\tablecaption{Percentiles of Morphology Classes}
\tablewidth{0pt}
\tablehead{
\colhead{Morph Class} & \colhead{Field} & \colhead{Cluster} & \colhead{Cluster } & \colhead{Cluster}\\
\colhead{ } & \colhead{ } & \colhead{Poor} & \colhead{All} & \colhead{Rich} \\
\colhead{$C$} & \colhead{(6051)} & \colhead{(346)} & \colhead{(3825)} & \colhead{(394)}
}
\startdata
0.35-0.50 & 21.6 & 20.8 & 28.6 & 30.0 \nl
0.25-0.35 & 45.2 & 43.1 & 44.9 & 48.1 \nl
0.10-0.25 & 33.2 & 36.1 & 26.5 & 22.0 \nl
\enddata
\end{deluxetable}

\end{document}